\documentclass[12pt]{article}
\usepackage{graphicx}

\newcommand{\be}{\begin{equation}}
\newcommand{\bea}{\begin{eqnarray}}
\newcommand{\ee}{\end{equation}}
\newcommand{\eea}{\end{eqnarray}}

\def\x{\mbox{{\bf x}}}

\def\R{\mbox{{\bf R}}}
\def\r{\mbox{{\bf r}}}
\def\S{\mbox{{\bf S}}}

\def\DF{\mbox{{\bf DF}}}

\def\eps{\mbox{$\epsilon$}}
\def\Y{\mbox{$\bf{Y}$}}

\def\X{\mbox{$\bf{X}$}}

\def\y{\mbox{$\bf{y}$}}

\def\u{\mbox{$\bf{u}$}}

\def\S{\mbox{$\bf{S}$}}

\graphicspath{%
    {converted_graphics/}
    {/}
    {E:/timedelayweather/jasonanfiles/D10Tau10g80/}
    {E:/timedelayweather/jasonanfiles/D10Tau10g50k80/}
    {E:/timedelayweather/jasonanfiles/0.51.5_data/}
    {C:/hdia/datassim/oceans/timedelayweather/jasonanfiles/datajason4042814/}
    {E:/timedelayweather/jasonanfiles/datajason4042814/}
    {C:/hdia/datassim/oceans/timedelayweather/jasonanfiles/datajason4042814/Datapackage4/D6_tau10/long/}
    {C:/hdia/datassim/oceans/timedelayweather/jasonanfiles/datajason4042814/Datapackage4/D8_tau10/}
    {C:/hdia/datassim/oceans/timedelayweather/jasonanfiles/datajason4042814/Datapackage4/D8_tau10/long/}
    {C:/hdia/datassim/oceans/timedelayweather/jasonanfiles/datajason4042814/Datapackage4/0.50.5/long/}
    {F:/timedelayweather/jasonanfiles/datajason4042814/Datapackage4/0.50.5/long/}
    {E:/timedelayweather/jasonanfiles/datajason4042814/Datapackage4/D8_tau10/long/}
    {E:/timedelayweather/jasonanfiles/datajason4042814/partiallymeasured/}
    {E:/timedelayweather/jasonanfiles/datajason4042814/Datapackage4/D6_tau10/long/}
    {E:/timedelayweather/jasonanfiles/datajason4042814/partiallymeasured/252/}
    {E:/timedelayweather/jasonanfiles/datajason4042814/Datapackage4/0.50.5/long/}
    {E:/timedelayweather/jasonanfiles/datajason4042814/partiallymeasured/248/}
}
\begin{document}

\begin{center}
{\bf {\large Estimating the State of a Geophysical System\\
with Sparse Observations: Predicting the Weather with an Accurate Initial State}}\\
\bigskip
\bigskip
Zhe An,\\
Daniel Rey,\\
Department of Physics,\\
\bigskip
and\\
\bigskip
Henry D. I. Abarbanel\\
Department of Physics\\
and\\
Marine Physical Laboratory (Scripps Institution of Oceanography),\\
\bigskip
University of California, San Diego\\
9500 Gilman Drive\\
La Jolla, CA 92093-0374,\\
\bigskip

\bigskip

\today
\end{center}
\newpage
\begin{abstract}
Utilizing the information in observations of a complex system to make accurate predictions through a quantitative model when observations are completed at time $T$, requires, even in the case where the model has no errors, an accurate estimate of the full state of the model at time $T$. In the case where the complex system has chaotic behavior this information transfer process, called data assimilation~\cite{cox,fried,even,apte,qjrms10,abar13}, encounters impediments to the estimations that are associated with the chaotic instabilities. These exhibit themselves as multiple local minima in a search for the state estimates, and these must be regularized to achieve accurate estimation~\cite{siads}. 

When the number of measurements $L$ at each observation time within the observation window is larger than a sufficient minimum value $L_s$, a property of the nonlinear dynamics of the model, the impediments in the estimation procedure are removed. As the number of available observations is typically such that $L \ll L_s$, additional information from the observations must be presented to the model. A striking example of wide interest is numerical weather prediction models which have order of 10$^8$-10$^9$ state variables and receive only 10$^7$ observations each 12 hours~\cite{ecmwf}.

We show how, using the time delays of the measurements at each observation time~\cite{abar13,rey14}, one can augment the information transferred from the data to the model, removing the impediments to accurate estimation and permitting dependable prediction. We do this in a core geophysical fluid dynamics model, the shallow water equations, at the heart of numerical weather prediction. The method is quite general, however, and can be utilized in the analysis of a broad spectrum of complex systems where measurements are sparse. When the model of the complex system has errors, the method still enables accurate estimation of the state of the model and thus evaluation of the model errors in a manner separated from uncertainties in the data assimilation procedure.

Our demonstration of this method for accurately initializing the state of a model to improve prediction has substantial potential to improve predictability with present and future models of the earth system, and especially for prediction of atmospheric processes. While our argument and methods are quite general, we have demonstrated them only in models with 10$^4$ state variables at most, and we recognize the challenges of a numerical implementation for models which are five orders of magnitude larger. Nonetheless, the power and generality of the method suggests it will provide an important missing ingredient in contemporary numerical weather prediction: accurate estimates in initializing the required model state.
\end{abstract}

\section{Introduction}

Prediction of the behavior of earth systems models lies at the core of numerical weather prediction, and given changes in system forcing over time, underlies our ability to forecast future short and long term behavior of our coupled ocean atmosphere system. Even if one should have a perfect model of the earth system, accurate prediction relies on the quality of our knowledge of the state of this model when predictions begin. Observations of the ocean/atmosphere/biosphere... are typically very sparse in that the number of state variables one observes, $L$, is much smaller than the number of dynamical variables $D$ of the model of the system: $L \ll D$. When the model system (and the observations) show chaotic behavior, there is a model dependent sufficient number of observations $L_s$ that permits accurate estimates of the state of the model when predictions are to commence. Typically, however, $L \ll L_s$ requiring one to find additional information from the observations to allow accurate estimations and dependable predictions. 

This paper explains how one can augment the information transferred from observations of a complex system using properties of the waveform of the observations in addition to point measurements at individual observation times. The method is demonstrated in calculations on a core geophysical model, the shallow water equations. The procedure of transferring information in observations to the estimation of parameters and states of a complex system, then testing the quality of the model through the metric of prediction, is called data assimilation, a name we adopt from the geophysical literature~\cite{even}. 

An important challenge in numerical weather prediction, or equivalent predictions in other complex systems, is the need to accurately initialize the model state at the time predictions commence. Our time delay method promises a substantial improvement in this initialization allowing increased accuracy in prediction when the model is accurate and significant increases in our ability to identify model errors through undependable predictions when the initialization of the model is adequate.

The discussion in this paper will focus on geophysical systems such as numerical weather prediction, while the methods we describe have broad applicability across the quantitative study of the underlying physical or biological properties in many complex systems. The features of having high dimensional dynamics and sparse $L \ll D$ measurements are typical in the process of examining the consistency of observed data and quantitative models of complex nonlinear systems, from functional nervous systems to genetic transcription dynamics to complex earth systems models and many other examples.

Why is this an issue at all? How can it be that presenting data on $L$ observed state variables to a $D$-dimensional model, even when $L < D$, cannot, through the use of the model as a nonlinear filter connecting the observed quantities to the unobserved states of the model, always allow one to adequately constrain and estimate the $D-L$ unobserved states? The answer to this lies in the chaotic behavior of a nonlinear model, of which weather models are contemporary and historical examples. 

Estimation procedures usually involve either (1) a minimization of a cost function acting as a metric for the closeness of observations and model output or (2) the sampling of a probability distribution for the state variables in the neighborhood of a maximum of the distribution. If $L < L_s$, the cost function has multiple local minima~\cite{siads} and in the second, the probability distribution has multiple maxima~\cite{abar13}, and these impede the  accurate estimation of the state variables at the end of observations. If the state of the model at the end of observations is inaccurate, then predictions will be undependable.

The paper examines the common situation where $L < L_s$ and explores how one may use time delayed samples of the observations to augment the information passed from the measurements to the model.

\section{Information in the Waveform of Observations}

\subsection{Transferring Information from Data to a Model}

We examine a setting where, during a temporal observation window $[0,T]$, we make $L$ observations $y_l(t_n);\;l=1,2,...,L$ at each time
$t_n = \{0=t_0,t_1,t_2,...,t_m = T\}$. Between observations the system moves its $D$-dimensional state $x_a(t);\;a=1,2,...,D$ ahead in time
via a set of ordinary differential equations
\be 
\frac{dx_a(t)}{dt} = F_a(\x(t)).
\label{eqmotion}
\ee 
If the dynamics of the system is described by partial differential equations, such as with the fluids in an earth systems model, we have realized them on a grid placing the grid label as well as the vectorial nature of the states into the index $a$. We have made the fixed parameters of the model into state variables with $F_a(\x(t)) = 0$ for them, and we have taken the dynamics to be deterministic. We can relax the discretization of space and work with stochastic dynamics (model errors), and we will indicate how to do the latter below. For illustration of the key idea, we retain the framework of Eq. (\ref{eqmotion}).

Within the idea that we know the model precisely, a good strategy is to adjust the unobserved states of the model from solving Eq. (\ref{eqmotion}) to minimize the least squares distance
\be 
\sum_{n=0}^m\biggl \{\sum_{l=1}^L\,(y_l(t_n) - x_l(t_n))^2\biggr\},
\label{cost1}
\ee
subject to the dynamical equations. 

Information can be presented to the model via the coupling or control (or in meteorology, nudging) achieved by adding to Eq. (\ref{eqmotion}) the terms
\be
\frac{dx_l(t)}{dt} = F_l(\x(t)) + \sum_{l'=1}^L\,g_{l,l'}(t)(y_{l'}(t) -x_{l'}(t));\; l=1,2,...,L \nonumber 
\label{control}
\ee 
for the observed states, $x_l(t);\;l=1,2,...,L$, and retaining the original equations
\be 
\frac{dx_q(t)}{dt} = F_q(\x(t)),\;q = L+1, L+2, ..., D, 
\ee
for the unobserved states. The $L \times L$ coupling matrix $g_{l,l'}(t)$ is positive definite. It is nonzero only in the neighborhood of a measurement time $t_n$. The role of the coupling term is to alter the Jacobian of the dynamical system (\ref{control}) allowing a suitable choice of $g_{l,l'}(t)$ to make all (conditional) Lyapunov exponents~\cite{pecora,abar96,kantz} negative, thus synchronizing the model output, for sufficiently strong coupling, with the observations $\y(t_n)$. We will see an explicit example of this later when we discuss the geophysical shallow water equation flow~\cite{ped,mwr13}.

Using Eq. (\ref{cost1}) is part of an approximation to the full answer to the data assimilation problem as given in~\cite{abar13}. We use this familiar approximation to motivate the key idea in the paper, and we discuss it within the general framework below.

One may minimize the cost function~(\ref{cost1}) to estimate $\x(t_0)$ and then integrate forward through the observation window to $t=T$ and beyond to predict, or using standard public domain numerical optimization packages one can estimate all $x_a(t_n)$ imposing Eq. (\ref{control}) as equality constraints between measurement times $t_n$. However, as we have shown in numerous examples, including the shallow water equation model we will discuss further, if $L$, the number of observations at each time $t_n$ is less than a sufficient number $L_s$~\cite{mwr13}, the search surface in the cost function will be filled with local minima, and this will impede the estimation of states, yielding inaccurate values for $\x(T)$ needed for prediction using Eq. (\ref{eqmotion}) for $t > T$. When $L \ge L_s$, this surface becomes smooth~\cite{abar13} and easy to search, giving accurate estimations of $\x(T)$ and good predictions for $t > T$.

In this paper we address how to proceed when $L < L_s$. 

\subsection{Adding Information from the Waveform of Measurements}

The idea is that information resides in the temporal derivatives of $y_l(t)$ in addition to that contained in the measurement itself at each $t_n$. As we do not have that derivative information directly, we may approximate it with known data as finite differences such as $[y_l(t_n + \tau) - y_l(t_n)]/\tau$, representing $dy_l(t)/dt$  with $\tau$ some multiple of the time differences between measurements~\cite{aeyels1,aeyels2,mane,embedology,takens,abar96,kantz}. As Takens noted in the context of nonlinear dynamical systems, the new information beyond $y_l(t_n)$ lies in $y_l(t_n + \tau)$, so that we can establish an extended state space by creating $L$ data vectors of dimension $D_M$ from $y_l(t_n)$ and its time delayed versions:
\be 
Y_{k;l}(t) = \biggl [y_l(t_n), y_l(t_n + \tau), ..., y_l(t_n+(k-1)\tau)\biggr];\;k=1,2,...,D_M.
\label{delay}
\ee 
This is a collection of $D_M$-dimensional time delays for each observed $y_l(t_n)$.

In nonlinear dynamics discussions  Takens' observation is employed to construct a proxy phase space of observations and their time delays which allows unambiguous orbits of the observed system by `unprojecting' from the projection associated with measuring only $L < D$ components of the underlying dynamics. 

Here our use of time delay coordinates is quite distinct. We only wish to choose $D_M$ large enough that the information in vectors such as in~(\ref{delay}) is enough to `cure' or regulate the lack of smoothness in cost functions such as Eq. (\ref{cost1}). This leads us to construct model vectors $S_{k;l}(\x(t))$ via the map
\be
\x(t) \to  S_{k;l}(\x(t));\;\;\;\; S_{k;l}(\x(t)) = \biggl [x_l(t), x_l(t + \tau),...,x_l(t+(k-1)\tau)\biggr],
\ee
with $k=1,2,...D_M;\; l=1,2,...,L.$

The dynamical rule in this space for the vectors $S_{k;l}(\x(t)$ is
\be
\frac{dS_{k;l}(\x(t))}{dt} = \sum_{a=1}^D \frac{\partial S_{k;l}(\x(t))}{\partial x_a(t)}F_a(\x(t))=\sum_{a=1}^D \frac{\partial S_{k;l}(\x(t))}{\partial x_a(t)}\frac{dx_a(t)}{dt}.
\ee

Following the idea of Eq. (\ref{control}), we introduce the control or regularization rule to present data to the model dynamics in this space of vectors $\S(\x(t))$ as
\be 
\frac{dS_{k;l}(\x(t))}{dt} = \sum_{a=1}^D \frac{\partial S_{k;l}(\x(t))}{\partial x_a(t)}F_a(\x(t)) + \sum_{k'=1}^{D_M} \sum_{l'=1}^{L} g_{k,l;k',l'}(t)\biggl\{Y_{k';l'}(t) - S_{k';l'}(\x(t))\biggr\},
\ee 
and then translate this regularized rule back into the physical space of the model variables $x_a(t)$ to arrive at 
\be 
\frac{dx_a(t)}{dt} = F_a(\x(t)) + g_{ab}\frac{\partial x_b(t)}{\partial S_{k;l}(\x(t))}g_{k,l;k',l'}(t)\biggl\{Y_{k';l'}(t) - S_{k';l'}(\x(t))\biggr\};
\label{control1}
\ee 
in which repeated indices are summed over. 

Two important differences are realized in Eq. (\ref{control1}) compared to Eq. (\ref{control}): (1) information from the time delays of the observations is presented to the physical model equations for $x_a(t)$, and (2) all components of the model state $\x(t)$ are influenced by the control term, not just the observed components, and the directions in phase space along which the controls are applied are determined by the inverse of the map $\x \to \S(\x)$. This means that parameters of the problem taken as state variables satisfying $F_a(\x(t)) = 0$ are also moved toward their values in the data.

Eq. (\ref{control1}) is our key result, specifying how information as realized in the time delay space of vectors $Y_{k;l}(\x(t))$ is transferred into the model, seen in the physical space $\x(t)$. The inverse of the rectangular $D \times (LD_M)$ dimensional matrix $\frac{\partial S_{k;l}(\x(t))}{\partial x_a(t)}$ is to be understood in terms of a generalized inverse~\cite{rey14}. When observations are available, they are coupled into the equations as shown. After observations, the model dynamics moves the system forward to the next measurement. 

Our procedure for determining the $D$-dimensional state $\x(t)$ within the observation window $t_0 = 0\le t \le T$ involves starting with an arbitrary initial condition $\x(t_0)$, perhaps using the known, observed state values for the components $x_l(t_0)$, and integrating the dynamical equation~(\ref{eqmotion}) forward to  $t_1 + (D_M -1)\tau$. Over the same interval $[t_0,t_1]$ we also integrate the variational equation for $\Phi_{ab}(t,t_0) = \partial x_a(t)/\partial x_b(t_0)$
\be 
\frac{d \Phi_{ab}(t,t_0)}{dt} = \sum_{c=1}^D \, DF(\x(t))_{ac}\Phi_{cb}(t,t_0);\;\;\;\;\Phi_{ab}(t_0,t_0) = \delta_{ab},
\ee
where $DF(\x)_{ab} = \partial F_a(\x)/\partial x_b$. 
At $t_1$, increment the computed value of $x_a(t_1)$ using $(\ref{eqmotion})$ by 
\be 
g_{ab}\frac{\partial x_b(t)}{\partial S_{k;l}(\x(t))}g_{k,l;k',l'}(t)\biggl\{Y_{k';l'}(t) - S_{k';l'}(\x(t))\bigr\}.
\ee
Entries of the required matrix $\partial S_{k;l}(\x(t))/\partial x_a(t)$ are elements of $\Phi_{ab}(t,t_0)$~\cite{rey14}.


Using the value $\x(t_1)$ as new initial conditions, integrate forward to $t_2 + (D_M-1)\tau$, and repeat the procedure to determine $\x(t_2)$. Continue this until one has utilized the information available at all of the measurements times $\{t_0,t_1,...,t_m = T\}$.

While doing these integrations, we monitor the success of the control terms in moving the model outputs $x_l(t)$ to the observed values $y_l(t_n)$
by calculating the synchronization error $SE(t)$ between them
\be 
SE(t)^2 = \frac{1}{L}\sum_{l=1}^L\,(x^{s}_l(t) - y^{s}_l(t))^2,
\label{syncherr}
\ee
in which we use the scaled variables $x^{s}_l(t) = (x_l(t)-x_{l-min}(t))/[x_{l-max}(t) - x_{l-min}(t)]$ where $x_{l-min/max}(t)$ is the minimum or maximum value of $x_l(t)$ up to time t. Similarly for $\y^{s}(t)$. This scales all data and observed model state values to lie in the interval $[0,1]$, and in this metric for synchronization treats the contribution of all state variables on an equal footing. As we integrate forward from $t_0$ the synchronization error will decrease if $L\,D_M$ is large enough indicating we have matched the model output more and more accurately. 

For $t > T$, we set $g_{k,l;k',l'}(t)$ and $g_{ab}$ to zero and integrate Eq. (\ref{eqmotion}) forward from the estimated $\x(T)$. This is the prediction window in which no information is passed from the data to the model and within which we can compare the model output $x_l(t > T)$ with further observations $y_l(t>T)$. The accuracy of these predictions is the metric for the quality of the model as well as a statement of the adequacy of the $L\,D_M$ measurements utilized in the state vectors $\S(\x(t))$ and $\Y(t)$.

We have found, and report now in some detail for an interesting geophysical model, that when the synchronization 
error Eq. (\ref{syncherr}) for a selected $L$, g, and $D_M$ decreases in time to very small values, the full state $\x(T)$ has been accurately estimated, and in simulations where we know the model generating the data precisely (called twin experiments), prediction is high quality. When the synchronization error for that selected $L$ and $D_M$ does not decrease to small values, $\x(T)$ is not well estimated, and prediction is undependable.

In practice, we have a given number of measurements $L$ at each observation time $t_n$. Using the model alone we can establish if this number suffices to move the model output to match the observed data, namely, $L \ge L_s$. If this is not the case and $L < L_s$, then by adding to the information transferred from the data to the model in a twin experiment, we can determine a $D_M$ which permits accurate estimations. Using the given value of $L$ and the determined value of $D_M$ to achieve small synchronization errors, we may use the model with the actual observed data to estimate values of the model state $\x(T)$ from which we can make predictions. 

The discussion has so far focused on a perfect model. As we have assured that the estimation process, namely the data assimilation procedure, with given $L$ and a given, perfect or imperfect, model and estimated $D_M$ will produce a good $\x(T)$, we can now predict for $t > T$ with the model knowing that failures in prediction arises from inadequacies in the model.
Once one has used this method to enhance and improve the model so that predictions are accurate, given additional observations we may use the model along with the methods described here to accurately estimate the state of the model from which we begin predictions.

\section{Shallow Water Equations}

To illustrate the implementation of the method of time delays utilizing the information latent in the waveform of a time series of observations, we describe how to proceed in detail in the shallow water flow on a $\beta$ plane. This geophysical fluid dynamical model~\cite{ped,mwr13} is at the core of earth systems atmosphere/ocean flows using in numerical weather prediction. Of course, production numerical weather prediction models contain much more detail than this core example and those models are meant to describe the dynamics over a whole sphere. We argue that the results we present for our simplified version of those massive models will be applicable, if numerically challenging, in establishing the state of those models after an observation window to allow adequate predictions and/or a focus on improving the model with assurance that the data assimilation procedure is satisfactory.

We perform what is called a `twin experiment'~\cite{twins1,twins2,twins3} in which we use a model to generate `data', then present $L$ measurements $y_l(t_n)$ from that data set to the model at each `measurement' time $t_n)$. The goal is to estimate the unknown parameters of the $D$-dimensional model and the $D - L$ unobserved states of the model using the subset of the measurements alone. This method tests the data assimilation technique, and, when successful, gives confidence in the method when applied to observed data. It also proves to be a way to expose the weaknesses and strengths of the estimation procedures, and, since we know all $D$ time series, to explore in detail how the unobserved variables are revealed by the procedure. 

As the depth of the atmosphere/ocean fluid layer (order 10-15 km) is markedly less than the earth's radius (6400 km), the shallow water equations for two dimensional flow are an excellent approximation to the fluid dynamics in such a geometry. Three fields on a mid-latitude ($\beta$) plane describe the fluid flow 
$(u(\r,t). v(\r,t), h(\r,t))$ with $\r = (x,y)$. These are the north-south velocity $v(\r,t)$, the east-west velocity $u(\r,t)$, and the height of the fluid $h(\r,t)$. The fluid is taken as a single constant density layer on a $\beta$-plane is driven by windstress $\tau(\r)$ at the surface $z = 0$ through an Ekman layer~\cite{ped,mwr13} yielding the dynamical equations
\bea
&&\frac{\partial \u(\r,t)}{\partial t} = - \u(\r,t) \cdot \nabla \u(\r,t) -g\nabla(h(\r,t) + \u(r,t) \times \hat{z}f(y) + \nu \nabla^2 \u(\r,t) - \epsilon \u(\r,t) \nonumber \\
&& \frac{\partial h(\r,t)}{\partial t} = -\nabla \cdot (h(\r,t) \u(\r,t)) - \hat{z}\cdot \mbox{curl} \biggl (\frac{\tau(\r,t)}{\rho f(y)}\biggr ).
\label{shallow}
\eea

Here $f(y) = f_0 +\beta y$, $\nu$ is the viscosity in the shallow water layer, and $\epsilon$ is the Rayleigh friction. We selected     $\tau(\r) = (F \cos(2\pi y), 0)$. $\hat{z}$ is a unit vector in the z-direction. The values we have used for the model parameters are given in Table 1. With these fixed parameters the shallow water flow is chaotic, and the largest Lyapunov exponent for this flow is $\lambda = 0.0325/hr \approx 1/31 hr$.

\begin{center}
\begin{table}
\begin{tabular}{|c|c|c|}
\hline
Parameter&Physical Quantity&Value in Twin Experiments\\
\hline
$\Delta t$&Time Step&36 $\mbox{sec}$\\
\hline
$\Delta X$&East-West Grid Spacing&50 $\mbox{km}$\\
\hline
$\Delta Y$&North-South Grid Spacing&50 $\mbox{km}$\\
\hline
$H_0$&Equilibrium Depth&5.1 $\mbox{km}$\\
\hline
$f_0$&Central value of the Coriolis parameter
&5 $\times 10^{-5}$ $\mbox{sec}^{-1}$\\
\hline
$\beta$&Meridional derivative of the Coriolis parameter
&2.0 $\times 10^{-11} (\mbox{sec-m})^{-1}$\\
\hline
$F$&Wind Stress&0.2 \mbox{m$^2$}/\mbox{sec$^3$}\\
\hline
$\nu$&Effective Viscosity&10$^{-4}$ \mbox{m$^2$}/\mbox{sec}\\
\hline
$\eps$&Rayleigh Friction&2 $\times 10^{-8}$ $\mbox{sec}^{-1}$\\
\hline
\end{tabular}
\caption{Parameters used in the generation of the shallow water `data' for the twin experiment. All fields as well as $(x,y,t)$ were scaled by the values in the table, so all calculations were done with dimensionless variables.}
\end{table}
\end{center}
We have analyzed this flow using the~\cite{Sadourny75} enstrophy conserving discretization scheme on a periodic grid of size $N^2$ with $N = 16, 32, 64$, and found that for the model with $D = 3 N^2$ degrees of freedom on this grid we must present approximately 70\% of D to achieve synchronization of the model output to the data thus determining the remaining state variables. We use the equations of motion Eq. (\ref{shallow}) to predict for times greater than the window $[0,T]$ in which we make observations. Of course, when the system is chaotic, small errors in the estimation of $\x(T)$, when observations finish, grow as $\exp[\lambda (t-T)]$.

These results on the required number of observations for the shallow water equations are in accord with calculations we presented earlier~\cite{mwr13}. As the results appear to be consistent for the various grid sizes we investigated, we present further results for the use of the time delay approach for N = 16. The total number of degrees of freedom is now 3N$^2$ = 768. For this we case establish this by examining synchronization between the data and model output in a twin 
experiment~\cite{twins1,twins2,twins3} that $L_s \approx 524 = 68\%$ of 3N$^2$. 

In~\cite{mwr13} we noted:
\begin{quotation}
``Attending to the number of required observations before carrying out a data assimilation procedure raises the difficult question of what one does when the number of observations is fewer than the number required for synchronization and accuracy in estimating the unobserved model state variables. We do not have an adequate answer to this question now."
\end{quotation}
In this paper we provide an answer to this key question. 

The question we address now is whether we can use the time delay method to nonetheless estimate $\x(T)$ accurately enough to predict for $t>T$ when $L \ll L_s$. We assume height measurements alone are made at each grid point $(i,j):\;i,j=1,2,...,N=16$, so $L = 256$. The velocity field $(u(i,j,t),v(i,j,t))$ is unobserved. We take all model parameters as known, though as indicated in the previous section one can estimate them in the same manner. Our focus is on accurately estimating the full state of the model at the end of observations when the number of observations at each observation time is $L < L_s$.

\begin{figure}[htbp] 
  \centering
  \includegraphics[bb=0 0 284 240,width=3.1in,height=3.6in,keepaspectratio]{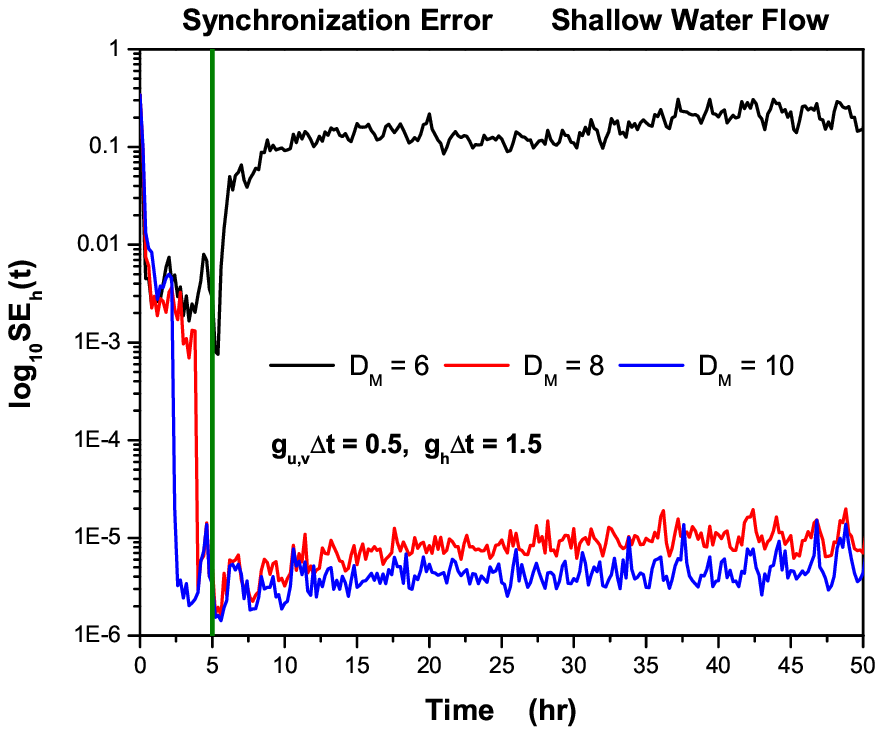}
 \includegraphics[bb=0 0 767 587,width=3.25in,height=3.8in,keepaspectratio]{hs64timedm6.eps}
  \includegraphics[bb=0 0 279 240,width=3.2in,height=3.7in,keepaspectratio]{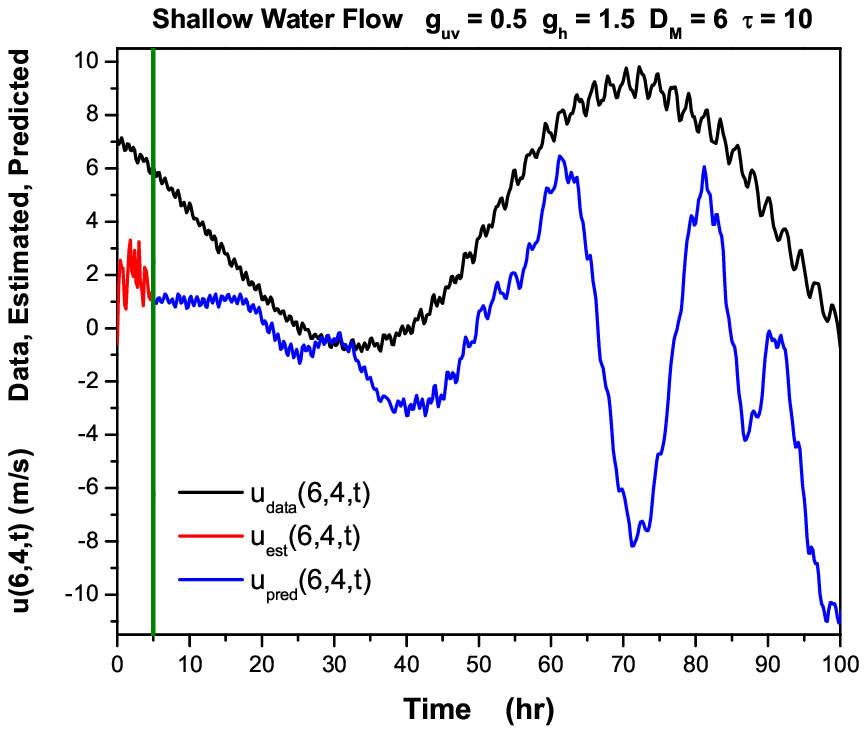}
  \includegraphics[bb=0 0 277 240,width=3.2in,height=3.7in,keepaspectratio]{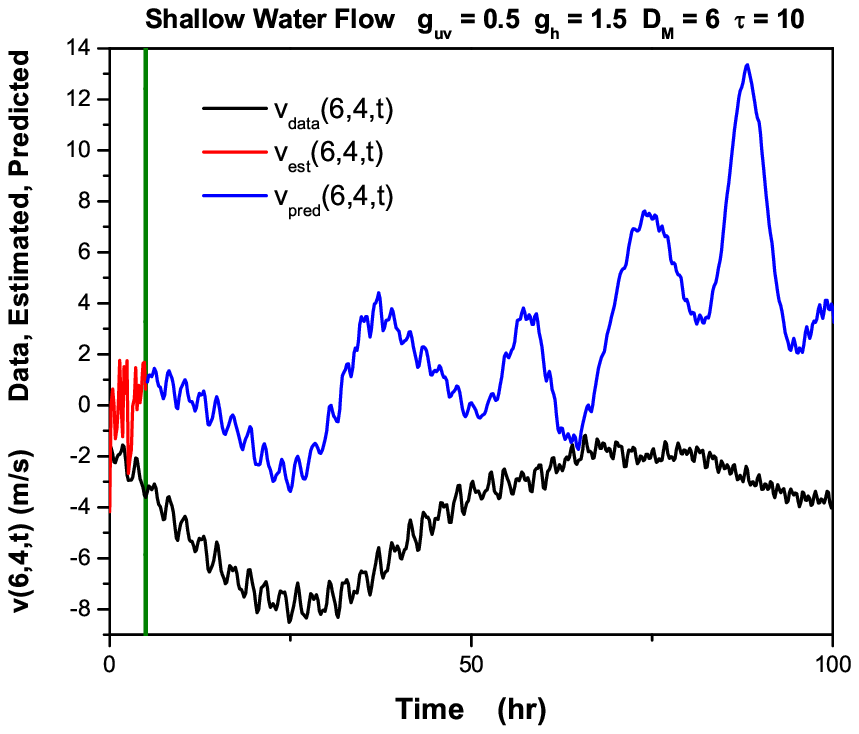}
  \caption{{\bf Upper Left} Synchronization error for the observed height field $h^s(i,j,t)$ (Eq. (\ref{syncherr}) for
couplings $g_{u,v}\Delta t = 0.5, g_h \Delta t = 1.5$, $\Delta t = 0.01\,$hr, $\tau = 10 \Delta t$ for time delay dimensions $D_M = \{6,8,10\}$.
{\bf Upper Right} Known (black), estimated (red) and predicted (blue) values for the scaled height field $h^s(i=6,j=4,t)$ when $D_M = 6$. The synchronization error and the scaled observed and predicted field are available in an actual experiment. In the twin experiment conducted here we also know the unobserved variables, the fluid velocities $(u(6,4,t),v(6,4,t))$ and these are displayed in the {\bf Lower } panels.}
  \label{sehdm6810tau10}
\end{figure}

In Fig. (\ref{sehdm6810tau10}) we display the synchronization error for the situation where we present L = 256 values of the height $h(i,j,t)$ on a 16 by 16 grid to the shallow water model using the modified dynamical equations with the control designated in Eq. (\ref{control1}) with coupling values of $g_{u,v} \Delta t = 0.5$ and $g_h \Delta t = 1.5$ with $\Delta t = 0.01\,$hr. We now take the coupling coefficients $g_{k,l;k',l'}(t)$ and $g_{ab}$ to be diagonal with different weights for the heights and for the velocities.

In the {\bf Upper Left} panel we show the synchronization error $SE_h(t)$ Eq. (\ref{syncherr}) for three selections of time delay dimensions $D_M = 6, 8, 10$ choosing the time delay to be $\tau = 10 \Delta t$. Our choice of time delay is a balance between numerical stability and the common nonlinear dynamical criterion of independent coordinates in a vector such as $\S(\x)$. The use of the first minimum of the average mutual information would select $\tau = 30 \Delta t$; we have chosen a smaller value which is still consistent with the standard convention.

The couplings are on for 500 time steps of $\Delta t$ or 5 hr, then they are removed. We see in this five hour observation window that for $D_M = 6$ the synchronization error starts and remains larger than order 0.005, then for $t > 5\,$ hr rises very rapidly to order 0.1. For $D_M = 8$ and $D_M = 10$, starting about the same magnitude falls rapidly within the observation window $0 \le t \le 5\,$hr to order 0.000005, and when the coupling is removed rises as $e^{\lambda (t-5)}$ with $\lambda$ very close to the calculated largest Lyapunov exponent for this flow. 

The quality of the synchronization, however, is then seen in the prediction for $t > T = 5\,$hr for the observed variables $h(i,j,t)$. In the {\bf Upper Right} panel of this figure we display with a black line the known height $h^s(6,4,t)$ at the grid point (i=6, j=4)  (chosen arbitrarily). We also show, with a red line, the estimated height $h^s(6,4,t)$ at that grid point for the observation window or estimation period, and for $t > T = 5\,$hr, the predicted value of $h^s(6,4,t)$ is in blue.

The prediction is not particularly good for $D_M = 6$, and we attribute this to a failure to accurately estimate the unobserved state variables, the fluid velocities u(i,j,t) and v(i,j,t)  during the observation period. In an actual experiment we would not be able to examine this statement as those velocities are not observed. However, this is a twin experiment, and in the {\bf Lower Left} and {\bf Lower Right} panels of Fig. (\ref{sehdm6810tau10}) we show the comparison between the known velocities, in black, at the grid point i=6, j=4 and, in red, the estimated velocities during the observations and the predicted velocities in blue after the observation window, namely, for $t > 5\,$hr. Indeed, we see that the estimates and the predictions are quite unacceptable for $D_M = 6$.

\begin{figure}[htbp] 
  \centering
\includegraphics[bb=0 0 767 587,width=4.2in,height=4.9in,keepaspectratio]{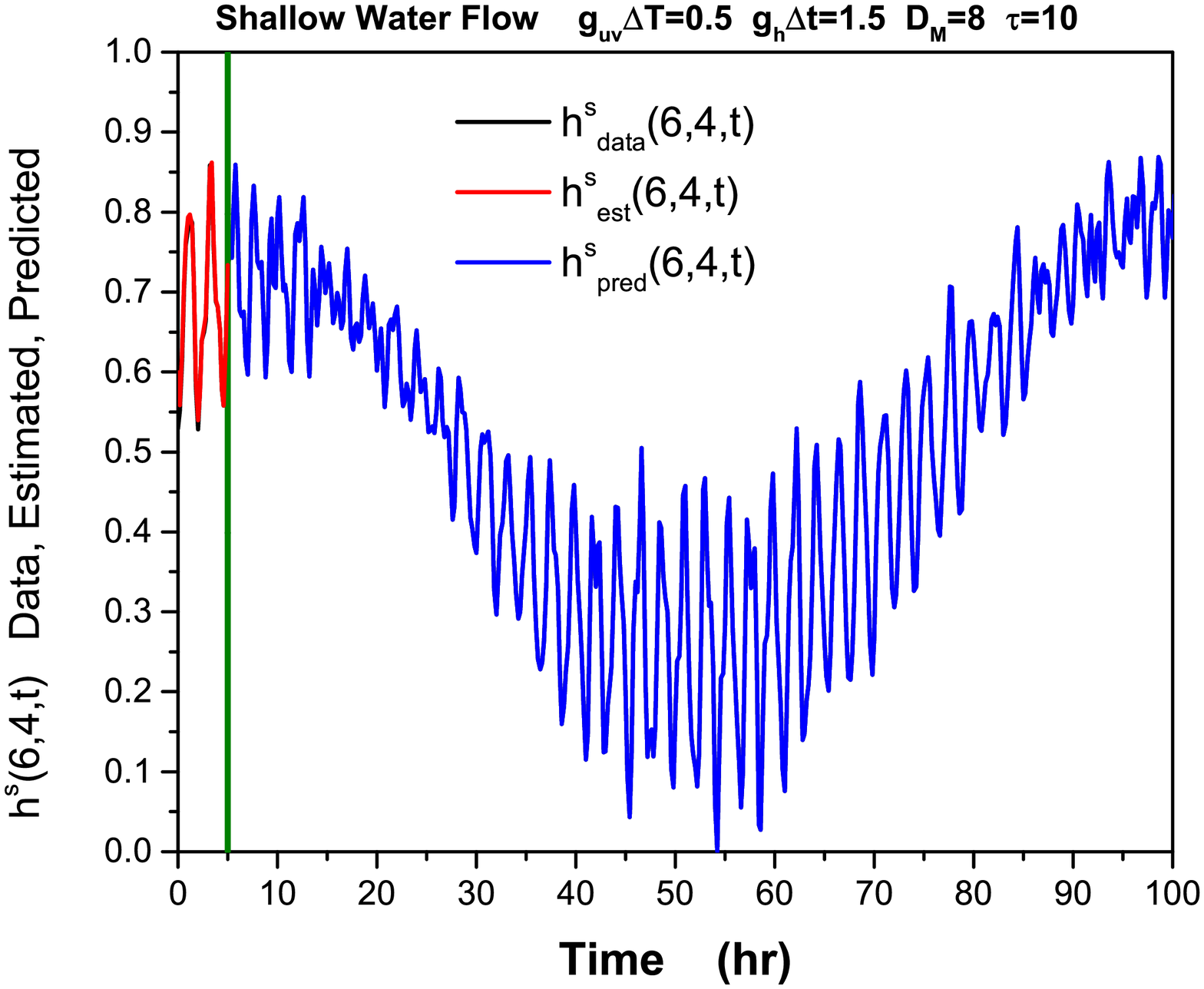}
  \includegraphics[bb=0 0 277 240,width=3.2in,height=3.7in,keepaspectratio]{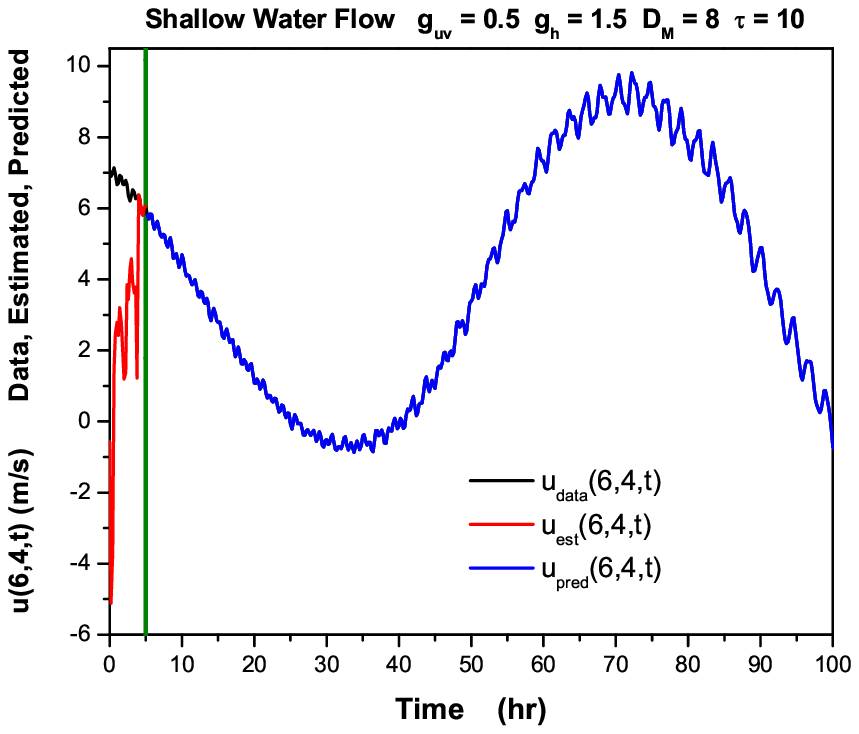}
\includegraphics[bb=0 0 275 240,width=3.2in,height=3.7in,keepaspectratio]{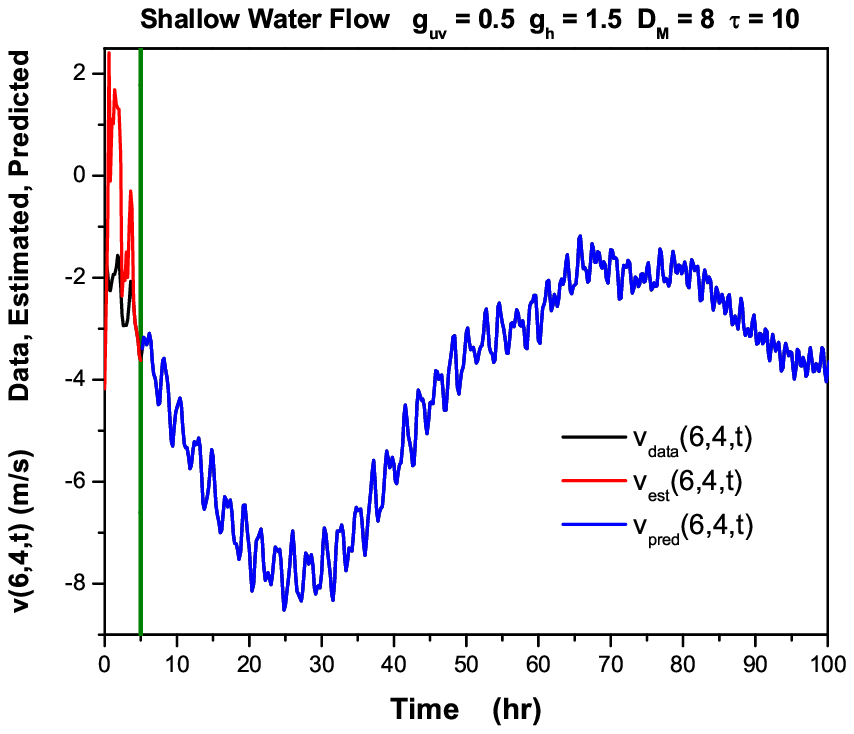}
  \caption{Same setting as for Fig. (\ref{sehdm6810tau10}) except the dimension of the time delay vectors is now $D_M = 8$. As the synchronization error $SE_h(t)$ is substantially reduced at $D_M = 8$ relative to what is seen with $D_M = 6$ we anticipate, and see, much more accurate estimation and prediction of the observable $h^s(i,j,t)$ and $\{u(i,j,t), v(i,j,t)\}$ fields for $D_M = 8$. {\bf Upper} panel: known (black), estimated (red), and predicted (blue) values for $h^s(6,4,t)$. {\bf Lower} panels: known (black), estimated (red), and predicted (blue) values for$\{u(6,4,t), v(6,4,t)\}$.}
  \label{uv64timedm8t10}
\end{figure}

In our display of the synchronization errors we saw that for $D_M = 8, 10$ $SE_h(t)$ was 10$^{-3}$ smaller than for $D_M = 6$. In Fig. (\ref{uv64timedm8t10}) we now display the three fields at grid location i = 6, j = 4 for $D_M = 8$ with all other choices of parameters held fixed. The {\bf Upper Panel} shows $h^s(6,4,t)$ for the known data in black, for the estimated ($0 \le t \le 5\,$hr, $h^s(6,4,t)$ in red, and for the predicted ($5 < t \le 100\,$hr in blue. The improvement in our predicting $h^s(6,4,t)$
is clear. The connection between the accuracy with which we know the fields at T = 5 hr and our ability to predict the observable $h^s(6,4,t)$ for $t > T$ is shown in this figure in the {\bf Bottom Left and Right} panels where the known, estimated and predicted unobserved velocities are shown. In an actual experiment we would not know the velocities as they are unobserved, however, here in a twin experiment we can examine in more detail the inner workings of the data assimilation procedure. The same striking improvement in estimation and prediction accuracy for all fields hold when we increase $D_M$ to 10; we do not display this.

In our modified dynamical equation Eq. (\ref{control1}) one role played by the control term is to change the original dynamics whose Jacobian matrix $\DF_{ab}(\x)$ leads to positive Lyapunov exponents associated with chaotic trajectories to a Jacobian dependent on the coupling strengths $g_{ab}$ whose conditional Lyapunov exponents are negative. The equation is conditioned on the presented data. This suggests that if we weaken the strengths with which the information in the data is coupled into the model dynamics, our estimation and prediction procedures may not work well. 

\begin{figure}[htbp] 
  \centering
  \includegraphics[bb=0 0 282 239,width=3.2in,height=3.7in,keepaspectratio]{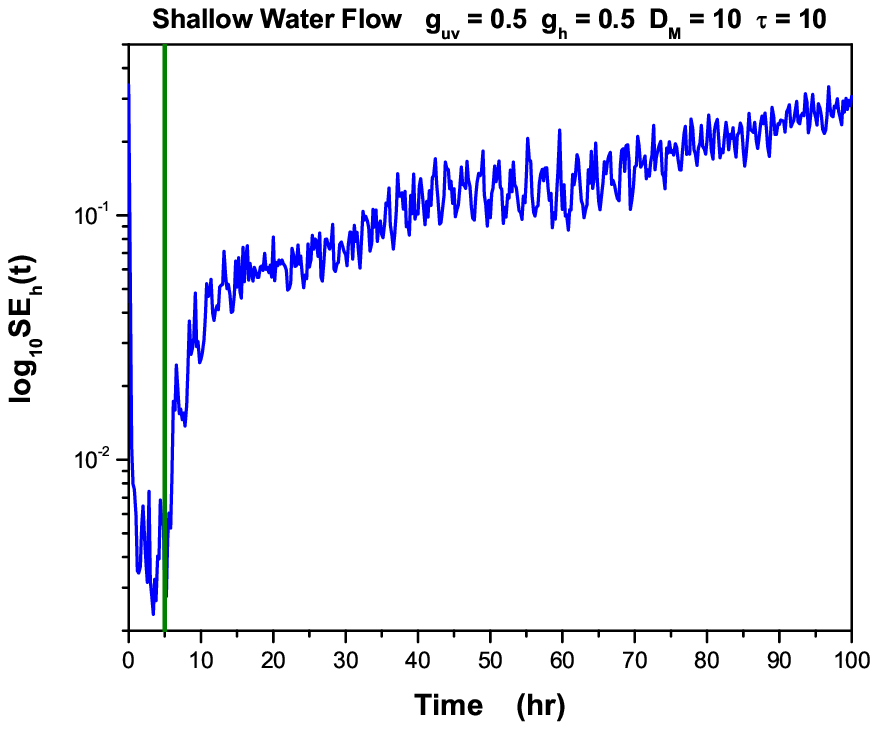}
  \includegraphics[bb=0 0 764 588,width=3.2in,height=3.7in,keepaspectratio]{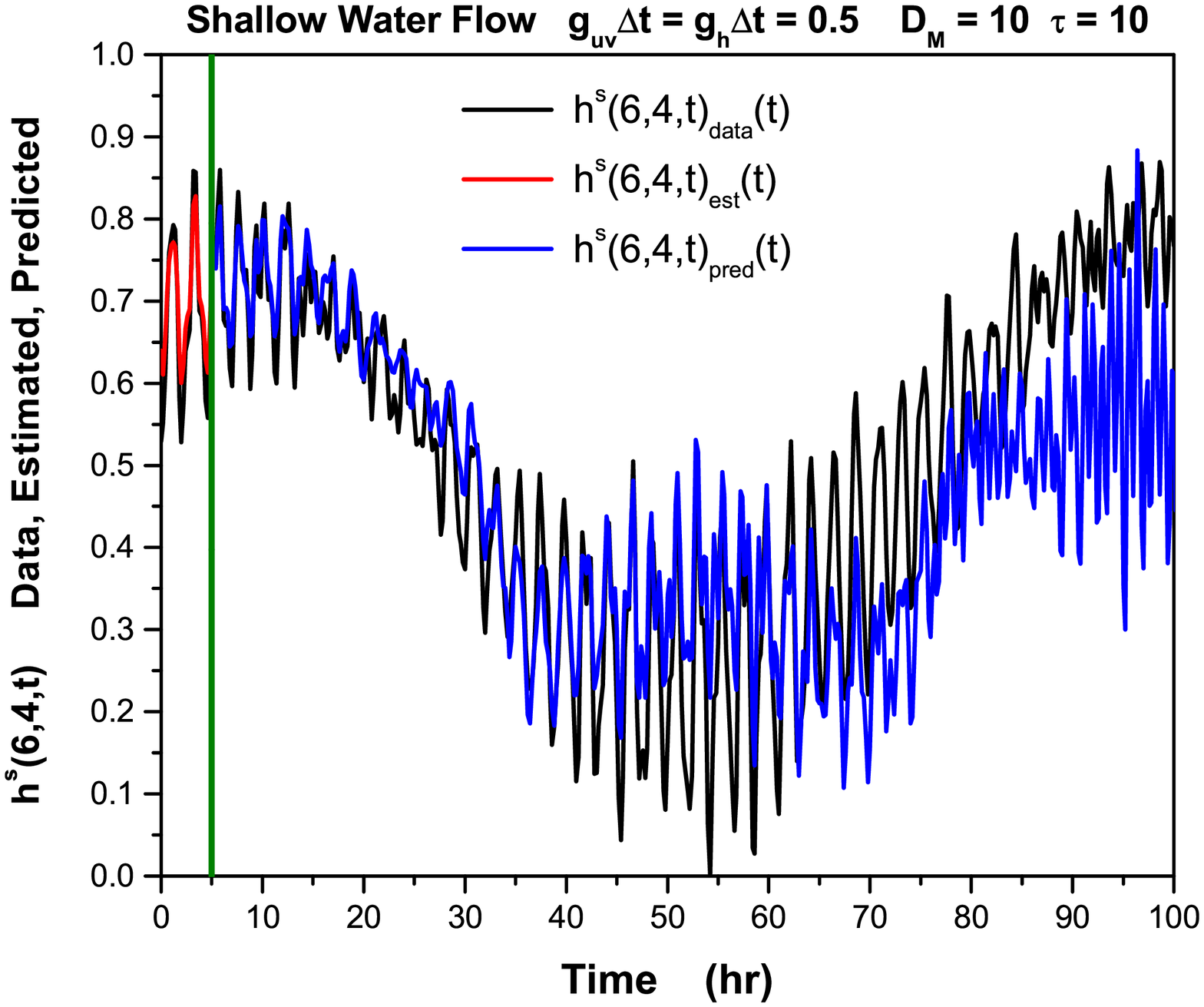}
  \includegraphics[bb=0 0 277 239,width=3.2in,height=3.7in,keepaspectratio]{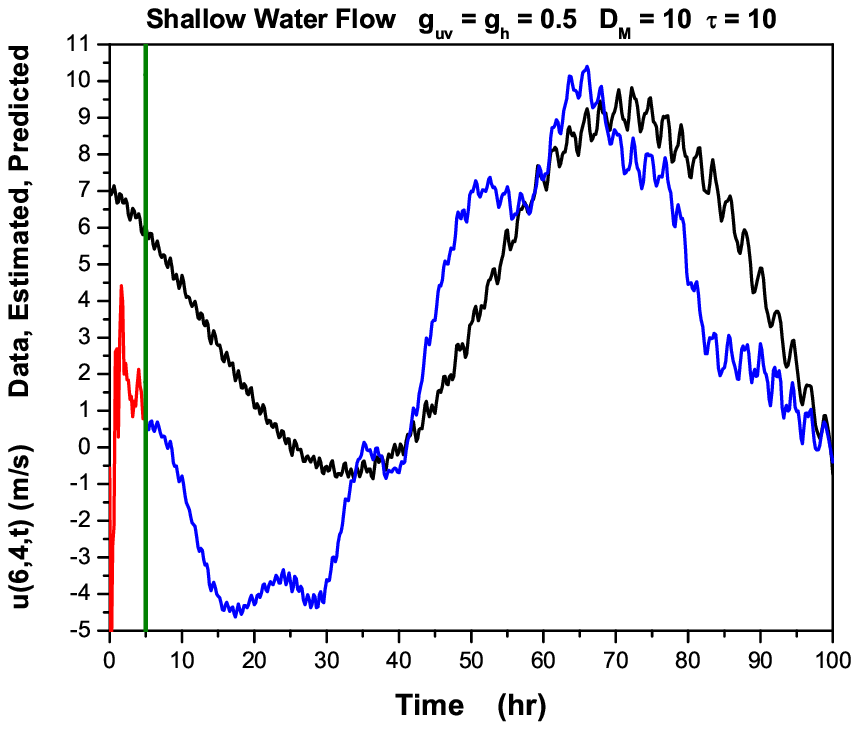}
  \includegraphics[bb=0 0 275 239,width=3.2in,height=3.7in,keepaspectratio]{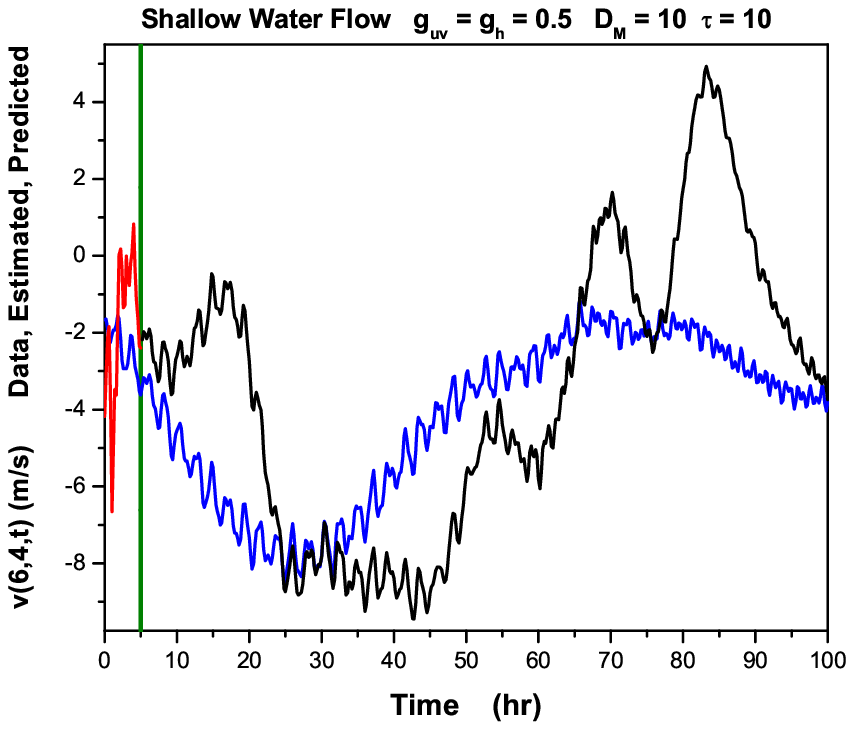}
  \caption{Caption for sehdm10t10g0.5}
  \label{sehdm10t10g0.5}
\end{figure}

We examine this in Fig. (\ref{sehdm10t10g0.5}) where we perform essentially the same calculations as reported just above, but we increase the number of time delays to $D_M = 10$ and reduce the coupling of the observations in the equation for $h(x,y,t)$ to $g_h \Delta t = 0.5$. In the {\bf Upper Left} panel we again display $SE_h(t)$ and see that we no longer arrive at the very well synchronized result we achieved with 
$g_h \Delta t = 1.5$. The absence of accurate synchronization of the model output from Eq. (\ref{control1}) is reflected in the {\bf Upper Right} panel where we display the
known, estimated and predicted values for $h^s(6,4,t)$ for this situation. In the lower panels we show the known, estimated and predicted values for the velocity at i = 6, j = 4 in this case of reduced couplings. Commensurate with what we have seen in the synchronization error and inaccurate values for the observable $h^s(6,4,t)$, we see that the velocities are not well estimated or predicted here. Just as a reminder we could not compare the estimated and predicted velocity fields in an actual experiment of this kind as they are not observed. In a twin experiment we can see rather more precisely what is happening.

\begin{figure}[tbp] 
  \centering
\includegraphics[bb=0 0 767 587,width=5.67in,height=4.34in,keepaspectratio]{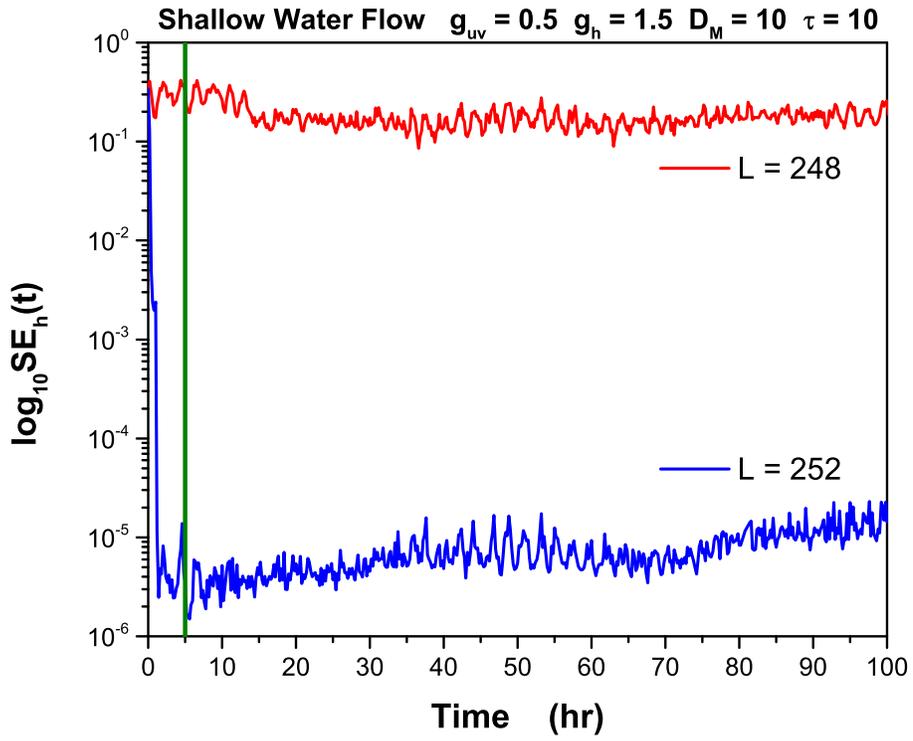}
  \includegraphics[bb=0 0 767 587,width=3.2in,height=3.7in,keepaspectratio]{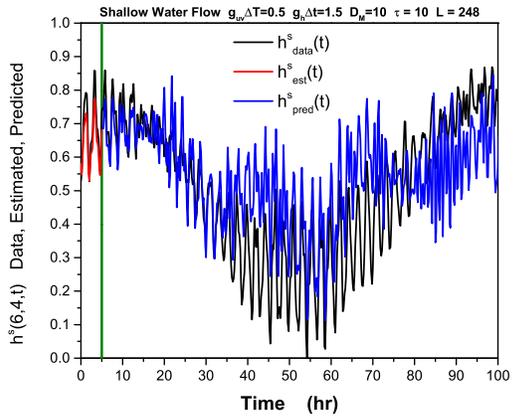}
  \includegraphics[bb=0 0 767 587,width=3.2in,height=3.7in,keepaspectratio]{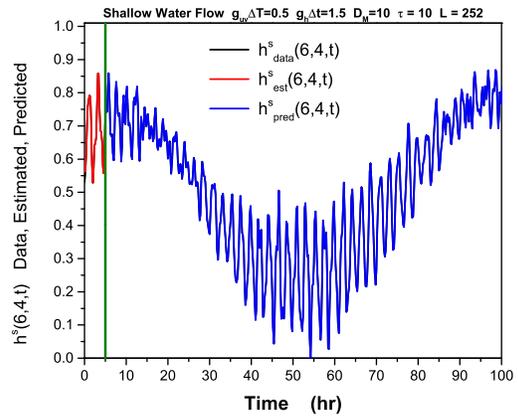}
  \caption{Caption for hs64timel252}
  \label{hs64timel252}
\end{figure}

In the calculations we have presented until now, we chose to observe the height field at all $L = N^2$ grid locations. We chose this value of $L$ for convenience, and now we ask if we can reduce it so as to place even fewer demands on the data collection. We found, as we shall now show, that reducing the value of $L$ very much, for these shallow water dynamics did not provide very much benefit. In Fig. (\ref{hs64timel252}) we display $SE_h(t)$ for $L = 252$ and for $L = 248$ which are rather close to the $L = 256$ we worked with above. All other parameters of the estimation procedure were kept fixed. 

Looking at the {\bf Upper} panel one can see that for $L = 252$ rapid and accurate synchronization is still achieved, while for $L = 248$ synchronization is basically not achieved at all. The {\bf Lower Left} panel then shows the known (black), estimated (red), and predicted values (blue) for $h^s(6,4,t)$ for $L = 248$. This confirms the message conveyed by $SE_h(t)$. In the {\bf Lower Right} panel we show the known (black), estimated (red), and predicted values (blue) for $h^s(6,4,t)$ for $L = 252$, and the connection between the synchronization error for this case and the ability to predict remains correct. Of course, one can display known, estimated and predicted versions of the unobserved fields, the velocity fields, as we have a twin experiment, but the figures here are what one would have in any real experiment, so we let this suffice here.

We have not thoroughly explored the overall parameter space of variables appearing in the use of time delayed measurements, namely, the couplings $g_{ab}$, the number of measurements $L$, the dimension of the time delay space $D_M$ and the time delay value itself $\tau$. It may very well be, for example, that adjusting all of these as we lower $L$ could provide excellent synchronization and prediction. 

We note again that the important issue we have addressed in this paper with the use of time delayed coordinates is that of initialization for prediction in complex systems. Our example of this within the shallow water equations provides a fruitful example of the approach which one can use to achieve accurate initialization, and should generalize to systems substantially larger than the one presented here.

The message we have tried to make clear with our selection of calculations is that using time delay measurements to transfer information to a model of observed 
dynamical processes allows one to achieve accurate state estimation and then prediction for high dimensional models for physical process of substantial interest. The core idea, couched in the examination of several `toy' models from nonlinear dynamics~\cite{rey14}, is available as well.

\section{Summary and Discussion}

The transfer of information from measurements on a chaotic complex system to a quantitative model of the processes in that system is impeded when the number of measurements at each measurement time is too small~\cite{siads,abar13}. The sufficient number of required measurements $L_s$ can be established by an examination of the model in a twin experiment, but the number of possible measurements $L$ may be such that $L < L_s$. This paper suggests, and explores in a key geophysical model of earth system flow, the shallow water equations, how one may augment further information from the measurements using the latent information in the waveform of the measured time series.
The additional information is embodied in the time delays of measurements at any given measurement time, and building measurement vectors from the point observations can overcome the impediments inherent in $L < L_s$.

We presented a data assimilation method which assumes no model errors, and we add control terms to permit the synchronization of data with the model output during a window of observations $[0,T]$. Using the estimated value of the $D$-dimensional ($D > L_s > L$) full model state $\x(T)$ we used the quality of the synchronization and, more importantly, the quality of the prediction for $t > T$ to assess the quality of the model. We varied the number of time delays, the number of measurements, and the strength of the couplings to demonstrate within the shallow water model how one may accurately initialize the model state preparatory to predictions using a sparse subset of observations.

In our discussion, focused on the use of time delay information, we took the model to be precise without model errors. This can be relaxed, and along with the natural errors in the observations and the uncertainty of the model state when observations begin, all cast into a statistical physics like path integral~\cite{abar13} within which one may answer all data assimilation questions, including the class of issues we addressed here. In that formulation one specifies an `action' $A_0(\X)$ on the path through the observation window $[0,T]$; $\X =\{x_a(t_0),x_a(t_1),...,x_a(t_m = T)\}; a=1,2,...,D$, and the probability density on the path $P(\X) = \exp[-A_0(\X)]$ determines the expected value of all functions of $\X$. The action is comprised of transition probabilities for the state $\x(t)$ to move to the state at the next time using a discrete time version of Eq. (\ref{eqmotion}) and a rule quantify the modification of the probability distribution associated with the transfer of information from noisy measurements the model with errors. The information transfer term in the action is essentially the cost function Eq. (\ref{cost1}). The model error, assuming Gaussian errors is proportional to
\be 
\sum_{n=0}^m \, \sum_{a,b=1}^D (\frac{dx_a(t_n)}{dt} -F_a(\x(t_n))\frac{\R_f(a,b)}{2} (\frac{dx_b(t_n)}{dt} -F_b(\x(t_n)),
\ee
all expressed in discrete time. $\R_f^{-1/2}$ indicates the reduced resolution in state space from the deterministic formulation.

Minimizing the action to determine the path is equivalent to the minimization of the synchronization error with model errors. This minimization process also has multiple minima which are removed by the addition of measurements, just as we have seen in the deterministic case discussed here for clarity~\cite{abar13}. 

To accommodate the use of time delay measurements, we can add to the action a term enforcing the map $\x \to \S(\x)$ locally and a term enforcing the variational equations in a stochastic manner. This then permits the estimation of a probability density for states during and following (in a prediction mode) the acquisition of observations. Instead of precisely known data and precisely known predictions as we show in our various figures, one has expected state values along with error bars on estimations and predictions~\cite{mwr13,abar13}.

This paper has presented both the idea of using time delay information to augment one's ability to accurately initialize a model state in order to predict and shown, in its application to geophysical shallow water flow with $D \approx 10^3-10^4$, how it can be implemented. We do not underestimate the numerical challenges involved in extending this to existing numerical weather prediction models with $D \approx 10^8-10^9$~\cite{ecmwf}. One idea would be to reduce the resolution of the model until one can use existing measurements to initialize the down scaled model, then interpolate to recapture the desired resolution for forward prediction. There sure are others including using the model to design additional observations suites to permit accurate initialization and prediction on the desired fine scale of the model.

The general method we propose permits the use of data assimilation in the analysis and understanding of complex dynamical systems with a focus on the properties and validity of the model itself. If the model is perfect with no model errors, we have shown how sparse measurements may be systematically augmented to permit accurate estimation and then prediction. Indeed, inaccurate estimation of $\x(T)$ is certain to lead to undependable prediction and errors in the understanding of the properties of the model even after assimilating information from data. If the model is wrong, as it typically will be in some or many aspects, the time delay method allows, with the use of twin experiments, the accurate estimation of the state of the model consistent with the available data, then prediction focuses on the verification of the model or the manner in which model improvements are to be accomplished without the concurrent worry that the procedure of estimation is masking the properties of the model.

\section*{Acknowledgements} We thank Dan Meiron for a critical discussion as we performed the calculations reported here. Detailed reading of the final draft of the paper by Massimo Vergassola and Lou Pecora is very much appreciated. This work was funded in part under a grant from the US National Science Foundation (PHY-0961153). Partial support from the Department of Energy CSGF program for D. Rey is appreciated. Partial support from the MURI Program (N00014-13-1-0205) sponsored by the Office of Naval Research is also acknowledged.

\end{document}